\documentclass{elsart}

\usepackage{natbib}

\usepackage{graphics}

\usepackage{amssymb}

\begin{document}

\begin{frontmatter}



\title{Measurement of the $D_{s}^{\pm}$ lifetime}

\centerline{(SELEX Collaboration)}


%

\author[Rome]{M.~Iori},
\author[PNPI]{G.~Alkhazov},
\author[PNPI]{A.G.~Atamantchouk\thanksref{tra}},
\author[ITEP]{M.Y.~Balatz\thanksref{tra}},
\author[PNPI]{N.F.~Bondar},
\author[Fermi]{P.S.~Cooper},
\author[Flint]{L.J.~Dauwe},
\author[ITEP]{G.V.~Davidenko},
\author[MPI]{U.~Dersch\thanksref{trb}},
\author[ITEP]{A.G.~Dolgolenko},
% ...
\address[Fermi]{Fermilab, Batavia, IL 60510, U.S.A.}
% ...
\address[ITEP]{Institute of Theoretical and Experimental Physics, Moscow, Russia}
\address[MPI]{Max-Planck-Institut f\"ur Kernphysik, 69117 Heidelberg, Germany}
% ...
\address[PNPI]{Petersburg Nuclear Physics Institute, St. Petersburg, Russia}
% ...
\address[Flint]{University of Michigan-Flint, Flint, MI 48502, U.S.A.}
\address[Rome]{University of Rome ``La Sapienza'' and INFN, Rome, Italy}
% ...
\thanks[tra]{Deceased}
\thanks[trb]{Present address: Infinion, M\"unchen, Germany}
\thanks[trc]{Now at Imperial College, London SW7 2BZ, U.K.}
\thanks[trd]{Now at Instituto de F\'{\i}sica da Universidade Estadual de Campinas, UNICAMP, SP, Brazil}
\thanks[tre]{Now at Physik-Department, Technische Universit\"at M\"unchen, 85748 Garching, Germany}
\thanks[trf]{Present address: The Boston Consulting Group, M\"unchen, Germany}
\thanks[trg]{Present address: Lucent Technologies, Naperville, IL}
\thanks[trh]{Now at Instituto de F\'{\i}sica Te\'orica da Universidade Estadual Paulista, S\~ao Paulo, Brazil}
\thanks[tri]{Present address: SPSS Inc., Chicago, IL}
\thanks[trj]{Now at University of Alabama at Birmingham, Birmingham, AL 35294}
\thanks[trk]{Present address: DOE, Germantown, MD}
\thanks[trl]{ Present address: Siemens Medizintechnik, Erlangen, Germany}
\thanks[trm]{Present address: Deutsche Bank AG, Eschborn, Germany}

\begin{abstract}
We report a precise measurement of the $D^{\pm}_{s}$ meson lifetime.
The data were taken by the SELEX experiment (E781) 
spectrometer using 600 GeV/$c$
$\Sigma ^{-}$, $\pi ^{-}$ and $p$ beams. 
The measurement has been done using 918 reconstructed 
$D^{\pm}_{s}$. The lifetime of the $D^{\pm}_{s}$ is measured to be
$472.5 \pm 17.2 \pm 6.6$ fs, using  $K^{*}(892)^{0}K ^{\pm}$ and  
$\phi \pi ^{\pm}$ decay modes. The lifetime ratio of $D_{s}^{\pm}$ to $D^{0}$ 
is $1.145 \pm 0.049$.
\end{abstract}


\end{frontmatter}

\input{psfig}

\section{Introduction}

Precise measurements of the lifetimes of charm meson weak decays
are important for understanding QCD  in both perturbative 
and nonperturbative regimes.  For mesons a joint expansion in Heavy Quark 
Effective Theory and perturbative QCD parameters 
treated through the third order in the heavy quark mass
shows a term including non-spectator W-annihilation 
as well as Pauli interference. The resulting non-leptonic decay rate
differences between 
W-exchange in $D^{0}$ and W-annihilation in $D_{s}^{\pm}$ produce 
lifetime differences of order 10-20 \% \cite{Frix97}.

The $D_{s}^{\pm}$ lifetime ~\cite{PDG} was dominated by the measurements
from E687 Collaboration (0.475 $\pm$ 0.020 $\pm$ 0.007 ps)~\cite{E687}.
 Recently new precision measurements of the $D_{s}^{\pm}$ lifetime have
been made by the E791 Collaboration 
(0.518 $\pm$ 0.014 $\pm$ 0.007 ps)~\cite{E791}
and the CLEO Collaboration (486.3 $\pm$ 15.0  $ ^{+4.9}_{-5.1}$ fs)~\cite{CLEO}.  Both groups have taken advantage of improved precision in the $D^0$
lifetime measurement to report new results for the 
$D_{s}^{\pm} to D^{0}$ lifetime ratio
of $ 1.254 \pm 0.041 $~\cite{E791} and $ 1.190 \pm 0.042 $~\cite{CLEO}. Their
average is 7.4 $\sigma$ from
unity, emphasizing the large difference in W contributions to 
$D_{s}^{\pm}$ and $D^{0}$ decays.

In this letter we report the results of a new measurement of the
$D_{s}^{\pm}$ lifetime based on data from the hadroproduction experiment
SELEX (E781) at Fermilab. The measurement is based on about 1000 fully reconstructed 
decays into $K K \pi $ from a sample of 15.3  $\times 10^9$ hadronic triggers.

The SELEX detector at Fermilab is a 3-stage magnetic spectrometer.
The negatively charged 600 GeV/$c$ beam contains nearly equal fractions of 
$\Sigma$ and $\pi$. The positive 
beam contains 92\% protons.
Beam particles are identified by a Transition Radiation detector.
The spectrometer was designed to study charm production in the forward 
hemisphere with good mass and decay vertex resolution for charm momenta
in the range 100-500 GeV/$c$. 
Five interaction targets (2 Cu and 3 C) had a total
target thickess of 4.2\% $\lambda _{int}$ for protons. The targets are
spaced by 1.5 cm. Downstream of the targets are 20  
silicon planes with a strip pitch of 20-25 $\mu$m oriented in X,~Y,~U and
V views.
The scattered-particle spectrometers have momentum cutoffs of 2.5 GeV/$c$ and 15 GeV/$c$ 
respectively. A Ring-Imaging Cerenkov detector (RICH)~\cite{rich}, filled with Neon at room temperature and 
pressure, provides single track ring radius resolution of 1.4\% and
 2$\sigma$ $K/ \pi$ separation up to about 165 GeV/$c$.
A layout of the spectrometer can be found elsewhere \cite{spec}.

\section{Data set and charm selection}

The charm trigger is very loose. 
It requires a valid beam track, at least 4 charged secondaries
in the forward 150 mrad cone, and two hodoscope hits after the second
bending magnet from tracks of charge opposite to that of the beam.
We triggered on about 1/3 of all inelastic interactions.
A computational filter linked PWC tracks having momenta $>15$ GeV/$c$ 
to hits in the vertex silicon and made a full reconstruction of primary and
secondary vertices in the event. Events consistent with only a primary vertex
are not saved. About 1/8  of all triggers are written to tape,
for a final sample of about $10^9$ events.

In the full analysis the vertex reconstruction was repeated with tracks of all
momenta. Again, only events inconsistent with having a single primary
vertex were considered. The RICH detector identified charged tracks above 25 
GeV/$c$. Results reported here come from a preliminary reconstruction through
the data, using a production code optimized for speed, not ultimate
efficiency. The simulated reconstruction efficiency of any charmed state
is constant at about 40\% for $x_{F}>0.3$ where $> 60\%$ of SELEX events lie.

To separate the signal from the noncharm background we require that: (i) the
spatial separation $L$ between the reconstructed production and decay 
vertices exceeds 8 times the combined error $\sigma _{L}$,  (ii)
each decay track, extrapolated to
the primary vertex $z$ position, must miss by a transverse distance
length $t \ge$ 2.5 times its error $\sigma _{t}$, (iii) the secondary
vertex must lie outside any target by at least 0.05 cm  and (iv) 
decays must occur within a fiducial region.

There are $918 \pm 53 $ events $D_s^{\pm}$ candidates, each having two 
RICH-identified kaons and a pion, for which no particle identification is
required. We divide them into three decay channels: $K^{*}(892)^{0}K ^{\pm}$, 
$\phi \pi ^{\pm}$ and other KK$\pi$. The resonant mass window for the 
$K^{*}(892)^{0}$ ($\phi$) was 
$892 \pm 70$  MeV/${c}^{2}$ ($1020 \pm 10$ MeV/${c}^{2}$).

$\pi /K$ misidentification causes a
reflection of $D^{\pm}$ under the $D_{s}^{\pm}$ peak. 
We limit the maximum kaon momentum to 160 GeV/$c$ to reduce 
misidentification in the RICH.  
To evaluate the shape of this background we use the  
$D^{\pm} \rightarrow K^{\mp} \pi^{\pm} \pi^{\pm}$  sample that passes 
all the cuts listed above and lies within $\pm$ 15 MeV/$c^{2}$ of the 
$D^+$ mass.  
We formed the invariant mass distribution of these  
events when one pion is interpreted as a kaon. At most one of the two possible
reflections per event falls into the $D_s^{\pm}$ mass window. The reflected 
mass distribution was fit by a polynomial function
rising at 1925 MeV/$c^{2}$ and decreasing to zero at large invariant mass.
Dividing this distribution by the number of $D^{\pm}$ events gives us the
contribution per mass bin for each misidentified $D^{\pm}$ in the $D_s^{\pm}$
sample. We count the misidentified $D^{\pm}$
in the $D_{s}^{\pm}$ sample by fitting the $D_s^{\pm}$ mass distribution 
within $\pm$ 20  MeV/$c^{2}$ interval around the $D_s^{\pm}$ mass with the 
sum of a Gaussian signal, a
linear background shape estimated from the sidebands and the $D^{\pm}$ shape
with variable normalization. The resultant misidentified $D^{\pm}$  
contribution to the $D_s^{\pm}$ mass distribution is shown
as the hatched areas in Fig. 1(a),~(b).  The fit gives $52 \pm 7$ and 
$12 \pm 3 $ misidentified $D^{\pm}$ events in the $K^{*}K$ and 
$\phi \pi$ decay mode, respectively (the error quoted is statistical only).
The RICH kaon identification is a very powerful tool for rejecting $D^{\pm}$
contamination; $\phi$ decay kinematics further reduces particle 
identification confusion in the $\phi\pi$ channel.    
 
To estimate yields, we subtracted the sideband background and $D^{\pm}$
contamination as evaluated above from the total number of events
in the signal region.  We find 430 $\pm$ 24
$K^*K$ and 330 $\pm$ 19 $\phi \pi$ events. A Gaussian fit to the combined 
data shown in Fig. 1(a),~(b) gives a mass of $1969 \pm 9.7 $ 
MeV$/c^{2}$ and $\sigma _{D_{s}^{\pm}}=8.0$ MeV$/c^{2}$. 
Note that these fitted yields are not
used as constraints in the lifetime fit, as discussed below.

\begin{figure}[h]
\centerline{\psfig{figure=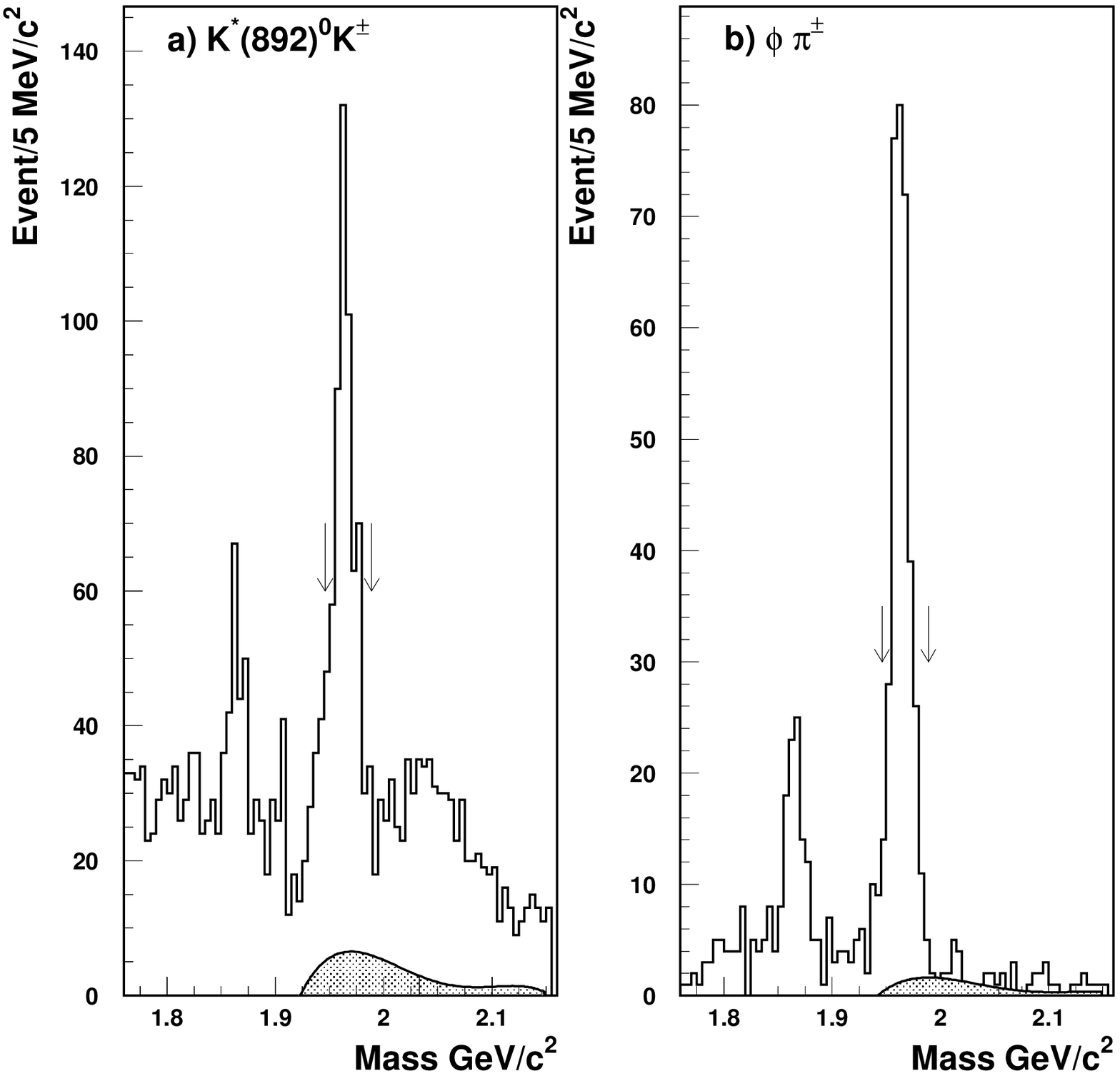}}
 
\caption{Invariant mass distributions for 
a) $D_{s}^{\pm} \rightarrow K^{*}(892)^{0} K^{\pm}$, 
b) $D_{s}^{\pm} \rightarrow \phi \pi ^{\pm}$.  
Hatched regions show the $K^{+} K^{-}$ $\pi^{\pm}$ background 
from misidentified $D^{\pm}$.  The arrows show the $D_{s}^{\pm}$ signal 
region.} 
\label{fig:1}
\end{figure}

 
\section{Lifetime evaluation using a maximum likelihood fit }

The average longitudinal error $\sigma _{z}$ on the primary
and secondary vertices is 270 $\mu$m and 500 $\mu$m, which gives
a combined error of 570 $\mu$m. In the $D_{s}^{\pm}$
sample, the average momentum is 215 GeV/$c$, 
corresponding to a time resolution of 18 fs, about $4 \%$ of 
$\tau _{D_{s}^{\pm}}$.
Because bin-smearing effects are small, we used a binned maximum likelihood
fitting technique to determine the $D_{s}^{\pm}$ lifetime.
The fit was applied to a reduced proper time distribution, 
$ t^{*} ={M(L- L_{min})/ p c }$ where
M is the known charm mass~\cite{PDG}, $p$ the reconstructed 
momentum, $L$ the measured vertex separation and
$L_{min}$ the minimum $L$ for each event to pass all the imposed selection 
cuts. $L_{min}$ is determined event-by-event, along with the acceptance,
by the procedure described below.  We fitted all events with 
$t^{*} < 1600$ fs in the mass range  
$ 1.949 < M(KK \pi ) < 1.989  $ GeV$/c^{2}$,
$ \pm ~2.5 \sigma$ from the $D_{s}^{\pm}$ central mass value. 

To evaluate the mean lifetime we used a maximum likelihood method.
The probability density was performed by the function :

$$f(\tau _{D_s},~\tau _{Bck},~\alpha,~\beta;~t^{*})=$$
$${(1-\alpha)N_{S}}{e^{-{t^{*}/{\tau _{D_{s}^{\pm}}}}}
\over{{\epsilon(t^{*}) \tau _{D_{s}^{\pm}}}}} + 
{\alpha}N_{S} B(t^{*}) 
+{N_{D^{\pm}}\over{\tau _{D^{\pm}}}}{e^{-{t^{*}/{\tau _{D^{\pm}}}}}}\eqno(1)$$
where
$$B(t^{*})={{\beta e^{-{t^{*}/{\tau _{Bck}}}}}\over{\tau _{Bck}}} +
{{1-\beta}\over{t^{*} _{Max}}}\eqno(2)$$

The function is the sum of a term for the $D_{s}^{\pm}$
exponential decay corrected by the acceptance function $\epsilon (t^{*}) $
 plus a background function $B(t^{*})$  consisting of a single exponential plus
 a constant to account for a
flat background extending to large proper time. Its parameters were
determined from the $t^{*}$ distribution from the $D_{s}^{\pm}$
sidebands.  It also includes
a term for the $D^{\pm}$ exponential decay normalized to the number of 
misidentified events in the signal region.  The $\tau _{D^{\pm}}$ lifetime
used in the fit is 1051 $\pm 13$ fs~\cite{PDG}. 
The mass range of the sideband
background windows,
 $1.890 < M(KK \pi ) < 1.930$ GeV$/c^{2}$   and
 $2.040 < M(KK \pi ) < 2.080$ GeV$/c^{2}$
was twice the signal mass window.
We defined asymmetric sidebands to avoid the influence of
$D^{\pm} \rightarrow K^{+}~K^{-}~\pi^{\pm}$, and we excluded the 
$D^{*}(2010)$ mass region.  

The four parameters are: $\tau _{D_{s}^{\pm}}$ ($D_{s}^{\pm}$ lifetime), 
$\tau _{Bck}$ (background lifetime), $\alpha$ (background fraction in the
signal region) and $\beta$ (background splitting function). 
$N_{S}$ is the total number of events in the signal region  
after $D^{\pm}$ contamination subtraction.

The proper-time-dependent acceptance  $\epsilon (t^{*}) $ is independent 
of spectrometer features after the first magnet, e.g., RICH efficiency and
tracking efficiency.  These efficiencies affect only the overall number of
events detected.  The proper time distribution of these events depends 
crucially on vertex reconstruction. To evaluate  $\epsilon (t^{*}) $ we 
reanalyze each $observed$ $D_s^{\pm}$ event after moving it to a large set of 
different proper
times $ t ^{*}$. Only the longitudinal position of the charm decay point
and the axial orientation of the 3-body decay vectors are changed 
\cite{sasha}.  
A reanalyzed event is
accepted if it passes the same cuts as those applied to the data.  The
minimum flight path after which an event is accepted defines $L_{min}$ for
this event.
 This method is independent of details of the true $x _{F} $ 
 or transverse momentum distributions \cite{sasha}.

Fig. 2 shows the overall fits to the data distributions as a function of 
reduced proper
time for $K^{*}(892)^{0}K^{\pm}$ and $ \phi \pi ^{\pm}$ decay modes. It also shows
the acceptance $\epsilon (t^{*})$, which differs
significantly from unity only after 4 lifetimes where statistics are limited.

Table 1 summarizes the lifetime results 
for the two modes analyzed:
$D_{s}^{\pm} \rightarrow K^{*}(892)^{0} K^{\pm}$; and
 $D_{s}^{\pm} \rightarrow \phi \pi^{\pm}$.
The uncertainties are statistical only, evaluated
where $ - \ln {\mathcal{L}}$ increases by 0.5.
Combining these results for the two resonant modes, 
we measure an average lifetime 
 $\tau _{D_{s}^{\pm}} = 472.5 \pm 17.2$ fs. 

\begin{figure}[h]
\centerline{\psfig{figure=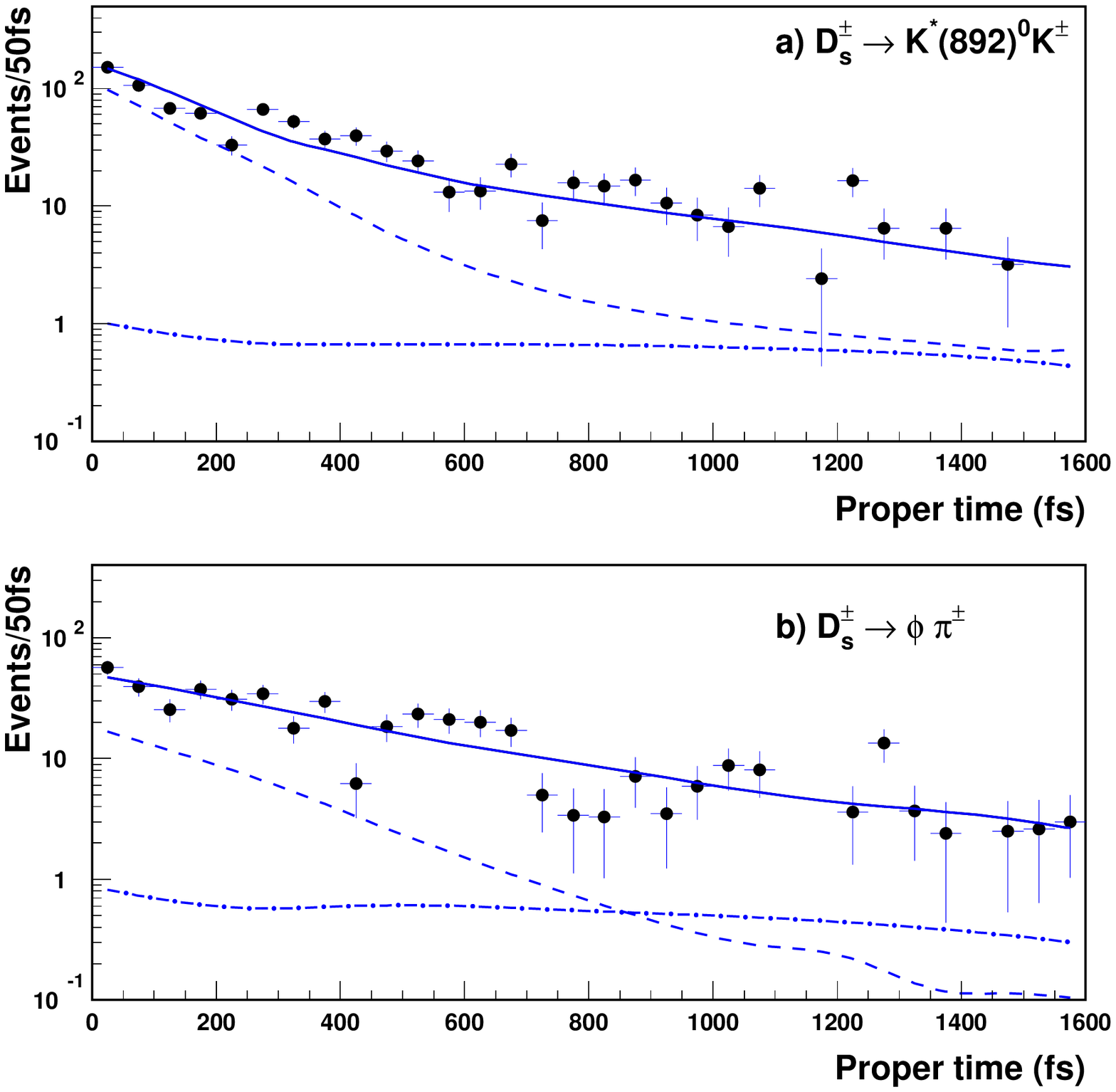}}

\caption{Corrected reduced proper time distributions for events in the 
$D_{s}^{\pm}$ window $1969 \pm 20$ Mev/$c^{2}$ (full circles) and results from the
maximum likelihood fit (solid curve) for:
a) ${D_{s}}^{\pm} \rightarrow {K}^{*}(892)^{0} K^{\pm}$; and b) ${D_{s}}^{\pm} \rightarrow \phi \pi^{\pm}$.
The dashed curve shows the fitted background, including 
the $D^{\pm}$ contribution. The dashed-dot curve describes the
acceptance function, $\epsilon(t^*)$.} 
\end{figure}


\begin{table}
\centering 
\begin{tabular}{|l|c|c|c|}  \hline\hline
  $D_{s}^{\pm}$  & $\tau _{D_{s}^{\pm}}$(fs)
 & Signal \\ \hline
 $ {K}^{*}(892)^{0}K^{\pm}$
& $472.3 \pm 23.0$  & $430 \pm 24$ \\ \hline\hline
  $ \phi \pi^{\pm}$ 
& $473.0 \pm 26.0$& $330 \pm 19$ \\ \hline\hline
 Average & $472.5 \pm 17.2$ & $760 \pm 30$ \\ \hline\hline
\end{tabular}
\caption{Lifetime results and signal yields for the two $D_s^{\pm}$ modes analyzed.
The last row is the weighted average of the two resonant channels
 $K^*(892)^{0}K^{\pm}$ and $\phi \pi^{\pm}$. The errors are statistical only.}
\end{table}

\begin{table} 
\label{sig}
\centering
\begin{tabular}{|l|c|c|} \hline\hline
Source of uncertainty&  $K^{*}(892)^{0} K^{\pm}$ &  $\phi\pi ^{\pm}$ \\ \hline\hline
Vertex reconstruction & $<$1 & $<$1 \\ \hline
$D ^{\pm}$ contamination & 2  & -- \\ \hline
Acceptance function &  3 & 2 \\ \hline
Fit procedure  &  5  & 0.3 \\ \hline
Total systematic error & 6.2 & 2.2 \\ \hline\hline
\end{tabular}
\caption{Systematic error contributions in fs.}
\end{table}

\section{Systematic errors}

The systematic uncertainties for the $D_{s}^{\pm}$ lifetime analysis are listed in Table 2 and described below. We group them in the following categories:

\subsection{Primary and secondary vertex reconstruction  }

Lifetime shifts due to reconstruction errors have been well studied in
our $D^0$ and $\Lambda_c$ work, with an order of magnitude higher
statistics \cite{sasha,E781}. Because of the high redundancy and good
precision of the silicon vertex detector, vertex mismeasurement effects are 
small at all momenta. Proper time assignment depends on correct momentum
determination. The SELEX momentum error is less than 0.5\% in all cases.  We 
assign a maximum systematic error from proper time
mismeasurement of 1 fs.

\subsection{Misidentification}

The effect of $D^{\pm}$ contamination under the $D_{s}^\pm$ peak was studied 
by changing the width of the exclusion window around the nominal $D^{\pm}$ mass
for the K/$\pi$ interchange discussed above. Effects on the $\phi \pi$ mode
are negligible.  For the $K^{*}(892)^{0}K ^{\pm}$ mode this gives a systematic error 
of 2 fs.  

\subsection{Acceptance function  }

The technique to determine the acceptance correction dependence
on proper time is discussed extensively in Ref.~\cite{sasha}.  It has
been verified with much larger statistics there.  The maximum systematic
error here is dominated by the $K^{*}(892)^{0}K ^{\pm}$ correction,
3 fs.  For the $\phi \pi ^{\pm}$ mode it is less than 2 fs.

\subsection{Fit procedure}
The fit was performed by the maximum likelihood method using a background
parametrized by an exponential function plus a constant. 
We varied the width of the sidebands and the $t^{*} _{Max}$ independently. 
The sytematic error due to the fit procedure is 5 fs
and less than 0.5 fs for $K^{*}(892)^{0}K ^{\pm}$ and 
$\phi \pi ^{\pm}$ decay modes respectively. That error is mainly dominated by the $D_s^{\pm}$ background parametrization.

Combining in quadrature all the sources of systematic errors listed in Table 2
we obtain a total systematic error of 6.2 fs (2.2 fs) for the 
$K^{*}(892)^{0} K ^{\pm}$  ($\phi \pi ^{\pm}$) mode.

\section{CONCLUSIONS}
We have made a new measurement of the $D_s^{\pm}$ lifetime
in two independent resonant decay channels, $K^{*}(892)^{0} K ^{\pm}$ 
and $\phi \pi ^{\pm}$ using a maximum likelihood fit.
SELEX measures the $D_s^{\pm}$ lifetime to be
$ 472.5 \pm 17.2 \pm 6.6$ fs.
Using the result reported in the PDG \cite{PDG}
$\tau _{D^{0}} = 412.6 \pm 2.8$ fs
we evaluate a ratio $\tau _{D_{s}^{\pm}}/{\tau_{D^{0}}} = 
1.145 \pm 0.049$, 3$\sigma$ from unity.  
The $D_{s}$ lifetime reported in this Letter is comparable
in precision with previous experiments \cite{E791,CLEO}. Our result,
combined with other world data,~\cite{PDG}, lowers the overall $D_s^{\pm}$ lifetime
somewhat.  Nevertheless, it is clear that the lifetime ratio
$\tau _{D_{s}^{\pm}}/{\tau_{D^{0}}}$ is significantly larger than unity
for all the precision measurements.

The authors are indebted to the staff of Fermi National Accelerator
Laboratory and for invaluable technical support from the staffs of 
collaborating institutions. This project was supported in part by
Bundesministerium f\"ur Bildung, Wissenchaft, Forschung und Tecnologie,
Consejo Nacional de Ciencia y Tecnologia (CONACyT), Conselho Nacional de
Desenvolvimento Cient\'ifico e Tecn\'ologico, Fondo de Apoyo a la
Investigaci\' on (UASLP), Funda\c c\~ ao de Amparo \'a 
Pesquisa do Estado 
de S\~ ao Paulo (FAPESP), the Israel Science Fundation founded by the
Israel Academy of Sciences and Humanities, Istituto Nazionale di Fisica
Nucleare (INFN), the International Science Foundation (ISF), the National Science Foundation (Phy 9602178), NATO (grant CR6.941058-1360-94), the Russian
Academy of Science, the Russian Ministry of Science and Technological Research
board (T\" UBITAK),the U.S. Department of Energy 
(DOE grant DE-FG02-91ER40664 and
DOE contract number DE-AC02-76CHO3000), and the U.S.-Israel Binational
Science Foundation (BSF).


\end{document}